\documentclass[letterpaper,twocolumn,pra,aps,10pt]{revtex4-1}
 \usepackage{mathptmx}
\usepackage{amsmath,amssymb,amsthm}
\newcommand\ket[1]{\,|#1\rangle}

\usepackage[colorlinks,citecolor=blue,linkcolor=blue,urlcolor=blue]{hyperref}
\begin{document}
\author{Michel Boyer$^{1}$,  Tal Mor$^2$\\
\small 1. D\'epartement IRO, Universit\'e de Montr\'eal,
  Montr\'eal (Qu\'ebec) H3C 3J7 \textsc{Canada} \\
\small 2. Computer Science Department, Technion, 
  Haifa 32000 \textsc{Israel}}
\title{Comment on ``Semiquantum-key distribution using less than four quantum states''}

\begin{abstract}
For several decades it was believed that information-secure key distribution
requires both the sender and receiver to have the ability to generate and/or 
manipulate quantum states. In 2007, we showed (together with Kenigsberg)
that quantum key distribution in which one party is classical, is possible.
A surprising and very nice extension of that result, was suggested   
by Zou, Qiu, Li, Wu, and Li. Their paper 
suggests that it is sufficient for the originator
of the states (the person holding the quantum technology) to generate just one
state! 

The resulting semiquantum key distribution, which we call here
``quantum key distribution with classical Alice'', is indeed completely robust 
against eavesdropping. 
However, their proof 
(that no eavesdropper can get information without being possibly detected), 
is faulty. We provide here a fully detailed and direct proof of their very important
result. 
\end{abstract}

\maketitle
A two-way Quantum Key Distribution (QKD) protocol 
in which one of the parties (Bob)
uses only classical operations was recently introduced~\cite{boyer:140501}. 
A very interesting extension in which the originator 
always sends the same state 
$\ket{+}$ (while in~\cite{boyer:140501} all four BB84 states are sent)
is suggested by Zou \textit{et al.}~\cite{zou:052312}. 
In both those semi-quantum-KD (SQKD) protocols 
the qubits  
go from the originator Alice to (classical) Bob and back to Alice.
Bob 
either reflects a received qubit without touching its state (CTRL), 
or measures it in the standard (classical) basis and sends back his 
result as $\ket{0}$ or $\ket{1}$ (SIFT).

We prefer to call 
the originator 
in~\cite{zou:052312}  
Bob (and not Alice), and to call the classical party Alice:
usually in quantum cryptography, Alice is
the sender of some non-trivial data, 
e.g., she is the one choosing the quantum states. 
The originator in~\cite{zou:052312} does not have that special role, 
as the state $\ket{+}$ is always sent 
(and we could even ask Eve to generate it). 
The classical person is then the one actually choosing a basis
and knowing which of the three state ($\ket{0}$, $\ket{1}$, or $\ket{+}$) is
sent back to the originator, thus it   
is natural to name that classical person Alice.
We call the originator Bob, 
and we call the SQKD protocol of Zou et al ``QKD with classical Alice''.

QKD with Classical Alice~\cite{zou:052312} is indeed completely
robust against eavesdropping. However, their proof 
(that no eavesdropper can get information without being possibly detected), 
is faulty.
Their result is stated in Theorem 5 which relies, after many steps,
on their Lemma 1.
Since the Lemmas are correct and only parts of the proofs are wrong, it is not at all 
easy to pinpoint errors. 
We explicitly single out two 
interwoven problems in Lemma 1:
1.--- They state (Lemma 1, first lines) that 
\textit{The originator's final state $\rho'^{B}$ is a product
state (in SQKD Protocol 2)}, prior to saying ``If the attack $(U_E,U_F)$
induces no error on CTRL...'', however, 
this statement is inaccurate;
in order to prove that Bob's final state is 
a product state one must make use of the fact that the attack induces no 
errors on CTRL  bits.
2.--- 
 They state 
(Proof of Lemma 1, first lines):
\textit{Because the originator (Bob) sends a qubit only after 
receiving the previous one, the qubits he received are in a tensor 
product form, i.e.  
${\rho'}^B = {\rho'_1}^B \otimes \cdots \otimes {\rho'_N}^B$.}
That is  not true in general. Consider, for instance, an attack on two 
consequative qubits: 
if Eve has a one qubit probe initialized as $\ket{0}$ and 
uses each of the two incoming qubits from Bob as a control bit
to apply a controlled-not gate (such that the NOT is applied onto her probe), 
then her probe keeps 
the parity of the two qubits sent by Bob; 
if
the classical party (Alice)  
reflects both qubits (CTRL),  the final global state is
$\frac{1}{2}\big[\left[ \ket{00}_{B} + \ket{11}_B\right]\ket{0}_E + \left[\ket{01}_B+\ket{10}_B\right] \ket{1}_E\big]$
and, once Eve's state is traced-out, the resulting state in 
Bob's hands 
is not a product state.

We now prove that the final result is indeed correct: robustness can be proven,
directly, as follows. The originator Bob keeps in a quantum
memory all qubits he received from Alice. When $N$ qubits have been
sent and received, classical Alice announces 
publicly the qubits she reflected; the originator Bob  
then checks that he received  $\ket{+}$
on those positions (CTRL).
For the qubits measured by Alice, a sample is chosen to be checked
for errors (TEST). 

Without loss of generality, we assume Eve uses a unique probe space for the attacks on all qubits and that her initial
state $\ket{E_0}$ is pure.
The analysis is now done bitwise, by induction. 
It is assumed that the Bob+Alice+Eve global state prior to Eve attacking qubit number $i$
is a tensor product state $\ket{\psi^{\,BA}_{i-1}}\otimes \ket{E_{i-1}}$ where $\ket{\psi^{\,BA}_{i-1}}$ is in Bob+Alice's hands
and Eve's current state $\ket{E_{i-1}}$ is independent of
all bits measured by Alice (SIFT).
That induction hypothesis obviously holds for $i=1$, before
the first qubit is sent~\cite{Note1}.
Eve knows that  Bob only sends $\ket{+}$ and she is free to send 
whatever state she wants to Alice. 
WLG (although she is assumed classical) Alice may delay measuring by
using a one-qubit probe and an XOR 
gate to SIFT~\cite{boyer:140501}.  The global state
before she decides whether she sifts or reflects can now 
be written 
\[
\ket{\psi^{\,BA}_{i-1}}\otimes \left[\ket{00}_{BA}\ket{E'_0} +\ket{10}_{BA}\ket{E'_1}\right]
\]
where $\ket{E'_b}$ are two un-normalized states of Eve's probe. In particular, if Eve ``does nothing'' ($\ket{E'_0}=\ket{E'_1}$),
Bob+Alices's state for the $i$-th qubit is $\ket{+0}$. 
On the qubit 
coming back, Eve applies the unitary $V_i$;
if Alice sifted, the global state before Eve applies $V_i$ is 
$\ket{\psi^{\,BA}_{i-1}}\otimes[\ket{00}_{BA}\ket{E'_0} +
  \ket{11}_{BA}\ket{E'_1}]$. 
Once Eve has applied $V_i$, it must be such that 
$V_i\ket{0}_B\ket{E'_0} = \ket{0}_B\ket{F'_0}$ else the TEST 
(in the classical basis) can detect an 
error, and similarly
$V_i\ket{1}_B\ket{E'_1} = \ket{1}_B\ket{F'_1}$. 
Due to the linearity of quantum mechanics, if classical Alice reflects
(CTRL), the 
resulting final state must be 
\[
\ket{\psi^{\,BA}_{i-1}}\otimes[\ket{00}_{BA}\ket{F'_0} + 
    \ket{10}_{BA}\ket{F'_1}].
\]
 Replacing now 
$\ket{0}_B$ by $[\ket{+}_B+\ket{-}_B]/\sqrt{2}$ and 
$\ket{1}_B$ by $[\ket{+}_B-\ket{-}_B]/\sqrt{2}$ 
gives
\[
\ket{\psi^{\,BA}_{i-1}}\otimes \left[
  \ket{+0}_{BA} \frac{\ket{F'_0} + \ket{F'_1}}{\sqrt{2}} + 
  \ket{-0}_{BA} \frac{\ket{F'_0} - \ket{F'_1}}{\sqrt{2}}\right];
\]
for  $\ket{-}_B$ to have probability $0$ of being measured by Bob, 
$\ket{F'_0} = \ket{F'_1}$ must hold; letting 
$\ket{E_i} = \sqrt{2}\ket{F'_0} = \sqrt{2}\ket{F'_1}$, the final global 
state is  
$\ket{\psi^{\,BA}_{i}}\otimes \ket{E_i}$ with 
$\ket{\psi^{\,BA}_{i}} = \ket{\psi^{\,BA}_{i-1}}\otimes\ket{\psi^{\,\prime BA}_{i}}$ 
where  $\ket{\psi^{\,\prime BA}_{i}} = (1/\sqrt{2})\left[\ket{00}_{BA} + \ket{11}_{BA}\right]$
if Alice shifts,
and $\ket{\psi^{\,\prime BA}_{i}} =\ket{+0}_{BA}$ if she reflects.

This completes the induction proof and
we deduce that after all $N$ qubits have been processed, the final global state is $\ket{\psi^{\, BA}_{N}}\otimes\ket{E_N}$ and 
Eve's state $\ket{E_N}$
is independent of all Alice's choices, and thus of her
information bits. That proves the robustness of the protocol. 

We proved here that the very nice protocol suggested by Zou et al (which 
we call here ``QKD with classical Alice'') is completely robust.
We would like to emphasize that the results
in~\cite{zou:052312} hold. As for Lemma 1, its statement can be slightly improved 
by moving the sentence ``Alice's final state ... is a product state...
in SQKD Protocol 2.'', till right after ``... the following conditions:".
Proving Lemma 1, however, is another matter, and it is unclear to us how
the original proof can be adjusted. Interestingly, it follows directly from our proof above that
the final Bob+Alice state  $\ket{\psi^{\,BA}_N}$ is  $\ket{\psi^{\,BA}_0}\otimes \bigotimes_{i=1}^N \ket{\psi^{\,\prime BA}_i}$,
where $\ket{\psi^{\,BA}_0}$ is the Bob+Alice state before the protocol, and the $\ket{\psi^{\,\prime BA}_i}$ are exactly those states announced by~\cite{zou:052312} in their Lemma 1,
which thus proves it for the ``one state'' protocol.
%
%

%


\end{document}